\RequirePackage{fix-cm}
\documentclass[smallextended]{svjour3}       
\smartqed  
\usepackage{graphicx}
\begin{document}

\title{Chiral Effective Field Theory after Thirty Years: \\ Nuclear Lattice Simulations
}


\author{Dean Lee}


\institute{Dean Lee \at
              Facility for Rare Isotope Beams and Department of Physics and Astronomy \\
              Michigan State University, East Lansing, MI 48824, USA \\
              \email{leed@frib.msu.edu}
}

\date{Received: date / Accepted: date}

\maketitle

\begin{abstract}
The introduction of chiral effective field theory by Steven Weinberg three decades ago has had a profound and lasting impact on nuclear physics.  This brief review explores the impact of Weinberg's work on the field of nuclear lattice simulations.  Rather than a summary of technical details, an effort is made to present the conceptual advances that made much of the recent progress possible.
\keywords{chiral effective field theory \and lattice simulations \and nuclear structure \and nuclear scattering and reactions \and nuclear thermodynamics}
\end{abstract}

\section{Introduction}
The author first learned of Weinberg's chiral effective field theory \cite{Weinberg:1990rz,Weinberg:1991um,Weinberg:1992yk} from the Loeb lectures that Weinberg delivered at Harvard in 1993.  While the author was a first-year graduate student at the time and didn't follow all the details, he appreciated the bold conclusion that there was an underlying hierarchy of nuclear forces.  Weinberg was showing which of the many possible nuclear forces were most important at low energies.  There was, however, some uncertainty if Weinberg's proposed hierarchy could withstand the combinatorial enhancement of higher-body forces in larger nuclei and many-body systems.  While the field has made great strides since then, this fundamental question still remains unresolved and is an active area of research today.  This article is a brief review dedicated to the memory of the inspirational physicist and pioneer we lost in 2021.

\section{Early lattice calculations}
The idea of combining lattice simulations with effective field theory is quite natural.  The lattice grid provides a short-distance regulator that can be used for an effective field theory.  The earliest lattice calculations of nuclear systems were posed in momentum space \cite{Brockmann:1992in}.  While this formulation of the lattice action makes the kinetic energy simple, it has the computational disadvantage that interaction terms require a summation over many momentum modes.  The earliest lattice calculations using a spatial lattice studied the equation of state of nuclear matter with single-site and nearest-neighbor interactions \cite{Muller:1999cp}.  Meanwhile, chiral perturbation theory for pions with lattice regularization was developed at around the same time \cite{Shushpanov:1998ms,Lewis:2000cc}, and there were also lattice formulations of chiral symmetry involving static baryons \cite{Chandrasekharan:2003wy}.

The first lattice calculations involving dynamical nucleons with chiral effective field theory appeared in Ref.~\cite{Lee:2004si}.  In that paper, pions were also treated as dynamical fields that coupled to the nucleon fields.  While pion-nucleon interactions produced the required pion-exchange interactions among the nucleons, they also altered the properties of individual nucleons such as the nucleon mass and the pion-nucleon coupling strength.  These interactions of the dynamical pions were problematic since they were being iterated to all orders without their corresponding short-distance counterterms.  Because of these complications, attention turned to simulations of pionless effective field theory, which didn't have such problems \cite{Lee:2004qd,Lee:2005is,Lee:2005it,Borasoy:2005yc}.  At the same time, several groups started applying lattice effective field theory to cold atoms and dilute neutron matter using pionless effective field theory \cite{Chen:2003vy,Bulgac:2005pj,Burovski:2006,Wingate:2006wy,Abe:2007fe}.  This review, however, focuses on nuclear lattice simulations using chiral effective field theory.

Later it was realized that the problems with the pions could be cured by removing the time derivative of the pion field from the lattice action \cite{Borasoy:2006qn}.  The resulting pion fields then only had spatial correlations without temporal correlations and produced the desired pion-exchange chiral potentials without modifying single nucleon properties and pion-nucleon vertices.  This first study considered chiral effective field theory at leading order.  The basic building block of the calculation is the Euclidean-time transfer matrix, $M$.  $M$ is defined as the normal-ordered exponential of the lattice Hamiltonian
$H$ over one time lattice step,
\begin{equation}
M = :\exp[-H\alpha_t]:, \label{Mtransfer}
\end{equation}
where $\alpha_t$ is the ratio of the time lattice spacing, $a_t$, to the spatial lattice spacing, $a$.  In the following, lattice units are used where all quantities are multiplied by powers of the spatial lattice spacing to make dimensionless combinations.  One considers an initial state $|\Psi_i\rangle $
and final state $|\Psi_f\rangle$
 that each have nonzero overlap with the nuclear state of interest.  By
multiplying by
powers of $M$ upon $|\Psi_i\rangle$, the
lowest energy state that couples to $|\Psi_i\rangle$ is enhanced.
The projection amplitudes take the form
\begin{equation}
Z_{f,i}(L_t) = \langle \Psi_f| M^{L_t}  |\Psi_i\rangle. 
\end{equation}
By calculating the ratio $Z_{f,i}(L_t)/Z_{f,i}(L_t-1)$ for large $L_t$, one can determine the energy of the lowest energy state.  By using more than one
initial and final state, a matrix of projection amplitudes can
be used to compute excited states as well.  

\section{Auxiliary fields and symmetries}

Auxiliary fields are used to generate the short-range nuclear interactions. The auxiliary field method can be viewed as a Gaussian integral formula connecting the exponential
of the two-particle density, $\rho^2$, to the integral of the exponential
of the one-particle density, $\rho$,\begin{equation}
{:\exp\left(-\frac{c\alpha_t}{2}\rho^2\right):}=\sqrt{\frac{1}{2\pi}}\int^{\infty}_{-\infty}ds
\, {:\exp \left(-\frac{1}{2}s^2 + \sqrt{-c\alpha_t}s\rho \right):}\;. \label{HS}
\end{equation}
The normal ordering symbols $::$ indicate placing all annihilation operators to the right and all creation operators to the left, along with the corresponding minus sign for each operator order interchange.  The normal ordering ensures that the operator products of the creation and annihilation operators behave as classical anticommuting Grassmann variables.  Auxiliary fields are introduced at every lattice site to produce the required two-particle interactions \cite{Hubbard:1959ub,Stratonovich:1958,Koonin:1986}.  Auxiliary fields can also be used to produce higher-body interactions \cite{Chen:2004rq,Korber:2017emn}. 
The pion fields are treated in a manner similar to the auxiliary
fields, except that the pion fields also have nontrivial correlations across
different spatial lattice sites.  

The amplitude for an $A$-body system is the determinant of an $A \times A$ matrix of single-nucleon amplitudes.  The resulting lattice calculations consist of computing matrix determinants for all possible values of the auxiliary fields.  These calculations are done using Markov Chain Monte Carlo sampling, and it is important that the determinant is not fluctuating strongly in sign or complex phase.  Fortunately, the fact that the two S-wave nucleon-nucleon channels have approximately equal strength gives rise to Wigner's approximate SU(4) symmetry \cite{Wigner:1936dx} of the low-energy nuclear interactions.  This approximate SU(4) symmetry for the attractive S-wave interactions suppresses the sign and phase oscillations of the matrix determinants.  This has been studied in several papers, some of which also consider inequalities that can be proven as a consequence \cite{Lee:2004ze,Lee:2004hc,Chen:2004rq,Lee:2007eu,Hoffman:2016jqv}.

The direct simulations of the higher-order terms in chiral effective field theory cause problems with sign and phase oscillations.  Some efforts were made to mitigate this problem using extrapolation methods \cite{Lahde:2015ona}, and others have taking a variational approach \cite{Wlazlowski:2014jna}.  But most of the efforts avoid this problem by computing high-order terms using perturbation theory \cite{Borasoy:2007vk,Epelbaum:2009rkz,Epelbaum:2009zsa,Epelbaum:2009pd,Epelbaum:2010xt,Lahde:2013uqa,Li:2018ymw}.  In order to determine all the unknown short-range operator coefficients, it is necessary to compute nucleon-nucleon phase shifts on the lattice.  While this can be done using Lüscher's finite volume method \cite{Luscher:1986pf}, the spherical wall method was found to provide greater accuracy for nucleon-nucleon scattering, especially for coupled partial-wave channels  \cite{Borasoy:2007vy,Lu:2015riz,Bovermann:2019jbt}.  In addition to higher-order corrections in chiral effective field theory, corrections needed to restore Galilean invariance have also been introduced \cite{Li:2019ldq}.  

Numerous calculations of nuclear structure and many-body systems have been performed, in particular the structure of alpha cluster states such as the Hoyle state in $^{12}$C \cite{Epelbaum:2011md,Epelbaum:2012qn,Epelbaum:2012iu,Shen:2021kqr} as well as states in $^{16}$O \cite{Epelbaum:2013paa}. 
As most other {\it ab initio} methods have difficulty computing such highly-correlated alpha cluster states, these works provide strong scientific justification for continuing efforts in lattice chiral effective field theory.  One important aspect of such calculations is the removal lattice artifacts from the final results.  Towards this end, there have been several developments to restore rotational symmetry \cite{Lu:2014xfa,Stellin:2018fkj} as well as remove finite-volume corrections from bound state calculations \cite{Bour:2011ef,Koenig:2011vaa,Davoudi:2011md,Korber:2015rce,Konig:2017krd}.

\section{Adiabatic projection method}

In order to compute nuclear scattering and reaction observables, a method called the adiabatic projection method was developed \cite{Pine:2013zja,Rokash:2015hra,Elhatisari:2016hby} and used to perform {\it ab initio} calculations of alpha-alpha scattering \cite{Elhatisari:2015iga}.  The adiabatic projection method constructs a low-energy effective theory for clusters of particles that is exact in the limit of large Euclidean time.
Consider a set of two cluster states labelled according to their spatial separation vector {\bf R}. The initial
wave functions are wave packets which, for large $|{\bf R}|$, factorize into a product of two individual clusters,
\begin{equation}
|{\bf R}\rangle=\sum_{{\bf r}} |{\bf r}+{\bf R}\rangle_1\otimes|{\bf r}\rangle_2.
\label{eqn:single_clusters}
\end{equation}
The summation over $\bf {r}$ produces states with
total momentum equal to zero. The initial cluster states are binned together
according to radial distance and angular momentum. This produces radial 
position states with angular momentum quantum numbers labelled as $|R\rangle^{J,J_z}$.

The next step is to multiply by powers of the transfer matrix in order to form ``dressed'' cluster
states that approximately span the set of low-energy cluster-cluster scattering
states in the periodic box of the lattice system. After $n_t$ time steps, the dressed cluster states are
\begin{equation}
\vert R\rangle^{J,J_z}_{n_t} = M^{n_t}|R\rangle^{J,J_z}.
\end{equation}
These dressed cluster states are then used to compute matrix
elements of the transfer matrix $M$,
\begin{equation}
\left[M_{n_t}\right]^{J,J_z}_{R',R} =\ ^{J,J_z}_{\!\!\quad{n_t}}\langle
R'\vert M \vert R\rangle^{J,J_z}_{n_t}.
\label{Hmatrix}
\end{equation}
Since such states are not orthogonal, it is also necessary to compute the norm matrix
\begin{equation}
\left[N_{n_t}\right]^{J,J_z}_{R',R} =\ ^{J,J_z}_{\!\!\quad{n_t}}\langle
R'\vert R\rangle^{J,J_z}_{n_t}. 
\label{eqn:norm}
\end{equation}
The ``radial adiabatic transfer matrix'' is defined as the matrix product
\begin{equation}
\left[ {M^a_{n_t}} \right]^{J,J_z}_{R',R} = 
\left[N_{n_t}^{-\frac{1}{2}}M_{n_t}
N_{n_t}^{-\frac{1}{2}} \right]^{J,J_z}_{R',R},
\label{eqn:Adiabatic-Hamiltonian}
\end{equation}
and a spherical hard wall boundary of radius $R_W^{}$ is imposed. For large $n_t$, the standing waves of the radial adiabatic transfer matrix can be used to determine the elastic phase shifts.  By including additional channels, the radial adiabatic transfer matrix can be used to compute inelastic reactions as well as capture reactions \cite{Rupak:2013aue}.

\section{Pinholes for structure and thermodynamics}

\begin{figure}[htb]
\centering
\includegraphics[width=9cm]{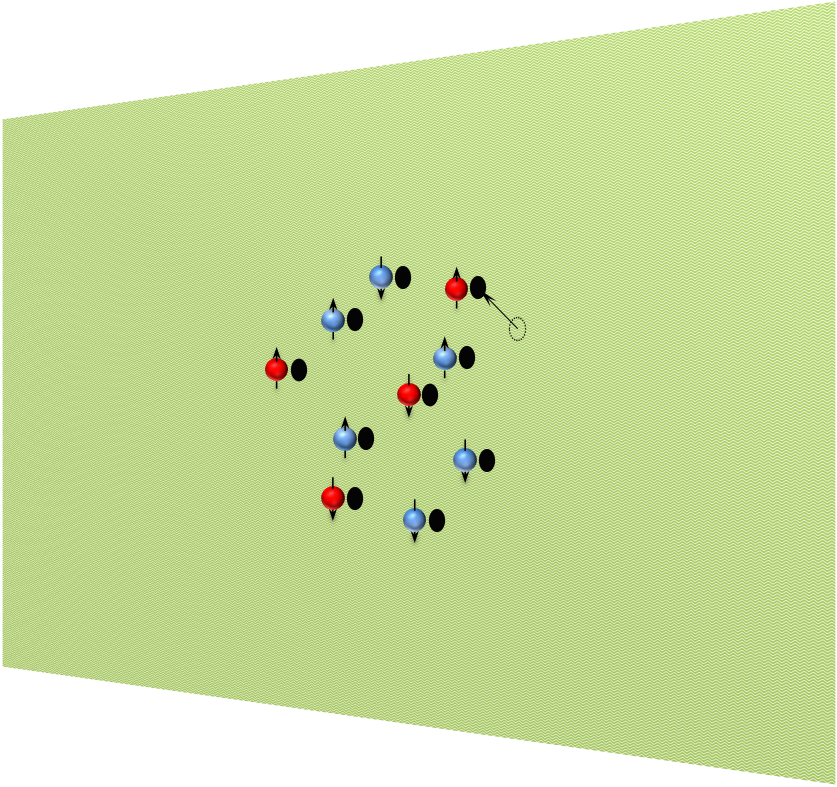}
\caption{Schematic drawing of nucleons and corresponding pinholes in the pinhole algorithm}
\label{pinhole}
\end{figure}

The amplitude for each auxiliary field configuration involves
quantum states which are superpositions of many different center-of-mass
positions.  Therefore, information about density correlations relative to
the center of mass is not straightforward to obtain. The pinhole algorithm solves this problem by computing a classical probability distribution for the positions of the nucleons \cite{Elhatisari:2017eno}.  Let $\rho_{i,j}({\bf n})$ be the density operator for nucleons with spin
$i$ and isospin $j$ at lattice site {\bf n},
\begin{equation}
\rho_{i,j}({\bf n}) = a^\dagger_{i,j}({\bf n})a_{i,j}({\bf n}).
\end{equation}
The normal-ordered $A$-body density operator is defined as
\begin{equation}
\rho_{i_1,j_1,\cdots i_A,j_A}({\bf n}_1,\cdots {\bf n}_A) = \; :\rho_{i_1,j_1}({\bf
n}_1)\cdots\rho_{i_A,j_A}({\bf
n}_A):.
\end{equation}
The $A$-body density operator is inserted in the middle of the Euclidean-time lattice simulations.  The amplitude will vanish unless there are $A$ nucleons that exactly match the spatial positions ${\bf n}_1,\cdots {\bf n}_A$ and spin-isospin indices $i_1,j_1,\cdots i_A,j_A$.   The spatial positions and spin-isospin indices can be viewed as ``pinholes'' in an otherwise impenetrable wall, and the pinhole positions and indices are updated using Monte Carlo sampling.  This is shown in Fig.~\ref{pinhole}.  The pinhole algorithm has been used to compute the density distributions of protons and neutrons in various nuclei \cite{Elhatisari:2017eno,Lu:2018bat}. 

In the area of nuclear thermodynamics, there have been recent lattice calculations of spin and density correlations of neutron matter at nonzero temperature \cite{Alexandru:2020zti}. A modified version of the pinhole algorithm called the pinhole trace algorithm was also developed for first principles calculations of nuclear thermodynamics \cite{Lu:2019nbg}.  In the pinhole trace algorithm, the trace over $A$-nucleon states is computed in position space,    
\begin{equation}
{\rm Tr}\; O = 
\frac{1}{A!} \sum_{i_1,j_1,{\bf n}_1 \cdots i_A,j_A,{\bf n}_A}
\langle 0 |
a_{i_A,j_A}({\bf n}_A) \cdots
a_{i_1,j_1}({\bf n}_1) \, 
O
\,
 a^{\dagger}_{i_1,j_1}({\bf n}_1) \cdots
a^{\dagger}_{i_A,j_A}({\bf n}_A)
| 0 \rangle.
\end{equation}
The pinhole trace algorithm was used to predict the liquid-vapor critical point in symmetric nuclear matter and probe alpha cluster formation as a function of temperature and density \cite{Lu:2019nbg}.  

\section{Nuclear forces and quantum phase transitions}

\begin{figure}[htb]
\centering
\includegraphics[width=11cm]{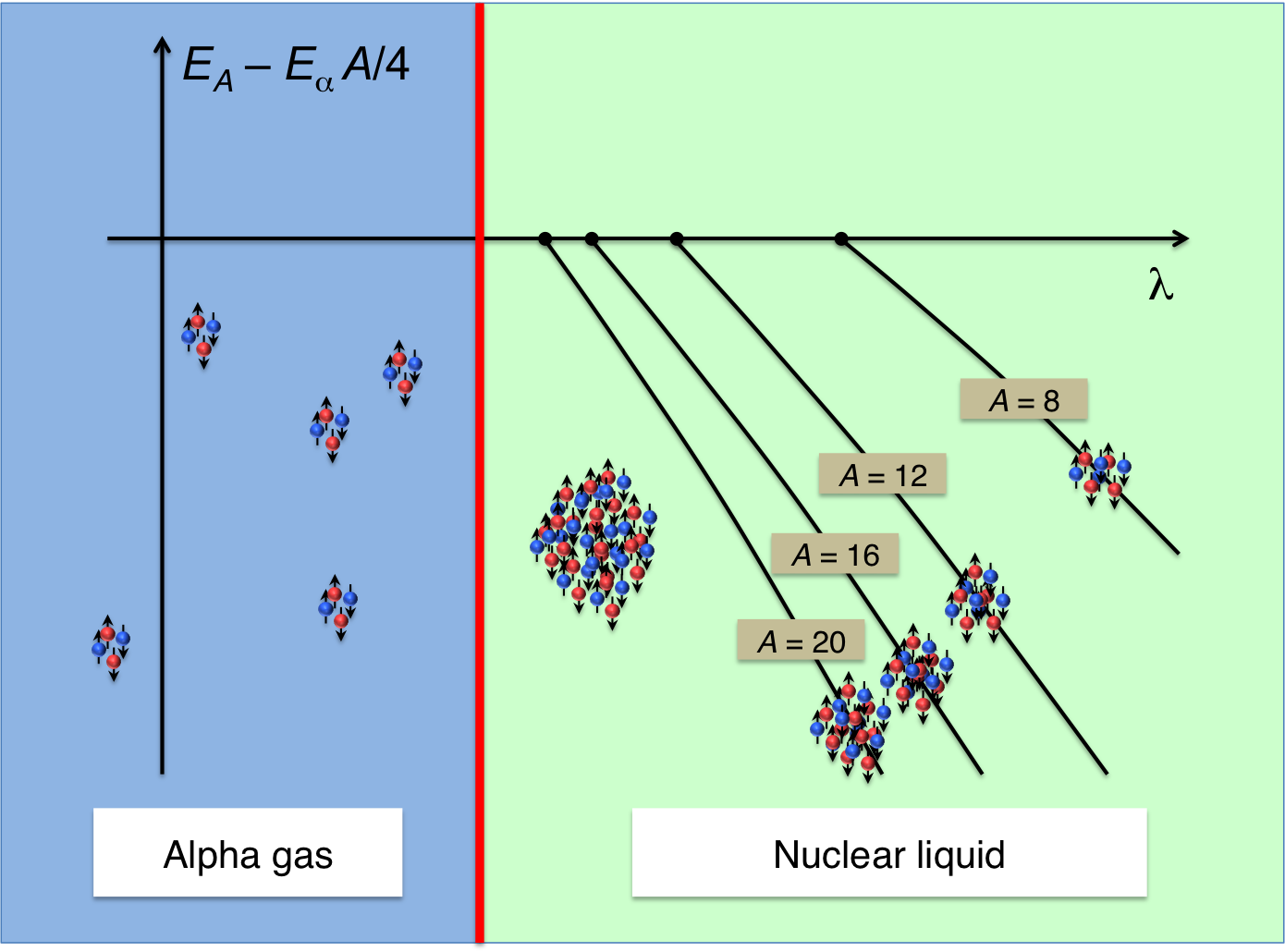}
\caption{Phase diagram of the ground state of symmetric nuclear matter as a function of the interpolating parameter $\lambda$ that controls the strength of the local part of the nuclear interactions.}
\label{phase_diagram}
\end{figure}

In addition to calculations using the most accurate lattice interactions available, there has also been a concerted effort to understand the most important characteristics of the nuclear force that determine the properties of nuclear binding.  In Ref.~\cite{Elhatisari:2016owd}, it was shown that nuclear physics resides near a quantum phase transition between a nuclear liquid and a Bose-Einstein condensate of alpha clusters.  By changing the range and locality of the nuclear force, one can traverse from one phase to the other.  The definition of locality here is that the nuclear interaction is diagonal with respect to particle position. In contrast, a nonlocal interaction is one that is off-diagonal with respect to nucleon positions.  Fig.~\ref{phase_diagram} shows the phase diagram of the ground state of symmetric nuclear matter as a function of a parameter $\lambda$ that controls the strength of the local part of the nuclear interactions.  The vertical axis is the energy relative to the corresponding multi-alpha-particle threshold.

The mechanism for this dependence on range and locality arises from a competition between attractive nuclear forces and the Pauli repulsion between identical nucleons \cite{Rokash:2016tqh,Kanada-Enyo:2020zzf}.  The radius of the alpha particle sets a characteristic length scale where the Pauli repulsion between identical nucleons in symmetric nuclear matter becomes strong.  The fact that nature appears to sit rather close to the quantum phase transition can be understood as arising from the phenomenological coincidence that the nucleon-nucleon interaction has an interaction range that is comparable to the radius of the alpha particle.
These findings were used to determine the simplest possible microscopic nuclear force that could accurately reproduce, on average, the ground state properties and charge radii of light and medium mass nuclei as well as neutron matter \cite{Lu:2018bat}.  The interaction was surprisingly simple and, except for the Coulomb interaction, respected Wigner's SU(4) symmetry.  This approximate SU(4) symmetry can be understood as resulting from an approximate spin-isospin exchange symmetry that appears when the number of colors in quantum chromodynamics is large  \cite{Kaplan:1995yg,Kaplan:1996rk,Banerjee:2001js,CalleCordon:2009ps,Timoteo:2011tt,RuizArriola:2016vap,Lee:2020esp}.  In Ref.~\cite{Lee:2020esp}, a derivation is given for the observations in Ref.~\cite{Timoteo:2011tt,RuizArriola:2016vap}
that there is an optimal resolution scale, corresponding to a momentum cutoff scale of about $500$~MeV, where spin-isospin symmetry is manifest.  These findings provide new insights into why chiral effective field theory performs remarkably well for momentum cutoff values near this scale.  They also provide new insights into the success of nuclear lattice simulations, which rely on the approximate spin-isospin symmetry to keep sign and phase oscillations small.

\section{Summary and Emerging Directions}
Detailed reviews of lattice effective field theory can be found in Ref.~\cite{Lee:2008fa,Drut:2012a,Lahde:2019npb}.  This brief review gives an overview of progress in nuclear lattice simulations using chiral effective field theory.  Nuclear lattice simulations is just one of many different methods to address the few-body and many-body problem in nuclear physics.  Perhaps its most important feature is that its strengths and weaknesses are very different from those of other approaches and thus provides valuable new information.  For example, nuclear lattice simulations is able to probe correlations in alpha cluster states and compute nuclear thermodynamics without much difficulty.    

This review started with the early history of nuclear lattice simulations, and it discussed the auxiliary field formalism and the importance of symmetries in keeping sign and phase oscillations in check.  The adiabatic projection method for nuclear scattering and reactions was reviewed, as well as the pinhole algorithm for structure and the pinhole trace algorithm for thermodynamics.  It also presented a discussion of the connection between nuclear forces and nuclear binding and the proximity of a quantum phase transition.

There are many new directions in nuclear lattice simulations now under active development.  The first lattice calculations using body-centered cubic lattices were recently completed \cite{Song:2021yst}.  Lattice calculations were able to determine the condensate fraction of a superfluid in the unitary limit \cite{He:2019ipt}, and the first lattice calculations of hypernuclei were performed \cite{Frame:2020mvv} using impurity lattice Monte Carlo \cite{Bour:2014bxa}.  The combination of lattice simulations with eigenvector continuation holds promise for calculating higher-order without relying on perturbation theory \cite{Frame:2017fah,Frame:2019jsw,Demol:2019yjt,Demol:2020mzd,Sarkar:2020mad}.  All of these developments, as well as others still in progress, show that the field of nuclear lattice simulations using chiral effective field theory is progressing and growing in relevance.  

\begin{acknowledgements}
The author is grateful to Jose Manuel Alarcón, Bu\=gra Borasoy, Lukas Bovermann, Jiunn-Wei Chen, Dechuan Du, Joaquín Drut, Serdar Elhatisari, Evgeny Epelbaum, Dillon Frame, Rongzheng He, Youngman Kim, Nico Klein, Sebastian K{\"o}nig, Hermann Krebs, Timo Lähde, Ning Li, Bing-Nan Lu, Yuanzhuo Ma, Ulf-G. Mei{\ss}ner, Michelle Pine, Alexander Rokash, Gautam Rupak, Avik Sarkar, Thomas Schäfer, Shihang Shen, Gianluca Stellin, Young-Ho Song, and all the members of the Nuclear Lattice Effective Field Theory Collaboration for their deep insights and productive collaboration.  Special thanks also to Ulf-G. Mei{\ss}ner for his careful reading and comments on this review.  Funding is gratefully acknowledged by the U.S. Department of Energy (DE-SC0013365 and DE-SC0021152) and the Nuclear Computational Low-Energy
Initiative (NUCLEI) SciDAC-4 project (DE-SC0018083) as well as computational resources provided by the Oak Ridge Leadership Computing Facility through the INCITE award ``Ab-initio nuclear structure and nuclear reactions'', the Gauss Centre for Supercomputing e.V.
(www.gauss-centre.eu) for computing time on the GCS Supercomputer JUWELS at J{\"u}lich Supercomputing Centre (JSC), the National Supercomputing Center at the Korea Institute for Science and Technology Information, and the high-performance computing centers at RWTH Aachen University and Michigan State University.
\end{acknowledgements}

\bibliographystyle{spmpsci}       
\bibliography{References.bib}   

\begin{thebibliography}{10}
\providecommand{\url}[1]{{#1}}
\providecommand{\urlprefix}{URL }
\expandafter\ifx\csname urlstyle\endcsname\relax
  \providecommand{\doi}[1]{DOI~\discretionary{}{}{}#1}\else
  \providecommand{\doi}{DOI~\discretionary{}{}{}\begingroup
  \urlstyle{rm}\Url}\fi

\bibitem{Abe:2007fe}
Abe, T., Seki, R.: Lattice calculation of thermal properties of low-density
  neutron matter with nn effective field theory.
\newblock Phys. Rev. \textbf{C79}, 054,002 (2009)

\bibitem{Alexandru:2020zti}
Alexandru, A., Bedaque, P., Berkowitz, E., Warrington, N.C.: {Structure Factors
  of Neutron Matter at Finite Temperature}.
\newblock Phys. Rev. Lett. \textbf{126}(13), 132,701 (2021).
\newblock \doi{10.1103/PhysRevLett.126.132701}

\bibitem{Banerjee:2001js}
Banerjee, M.K., Cohen, T.D., Gelman, B.A.: {The Nucleon nucleon interaction and
  large N(c) QCD}.
\newblock Phys. Rev. C \textbf{65}, 034,011 (2002).
\newblock \doi{10.1103/PhysRevC.65.034011}

\bibitem{Borasoy:2006qn}
Borasoy, B., Epelbaum, E., Krebs, H., Lee, D., Mei{\ss}ner, U.G.: {Lattice
  Simulations for Light Nuclei: Chiral Effective Field Theory at Leading
  Order}.
\newblock Eur. Phys. J. A \textbf{31}, 105--123 (2007).
\newblock \doi{10.1140/epja/i2006-10154-1}

\bibitem{Borasoy:2007vy}
Borasoy, B., Epelbaum, E., Krebs, H., Lee, D., Mei{\ss}ner, U.G.: {Two-particle
  scattering on the lattice: Phase shifts, spin-orbit coupling, and mixing
  angles}.
\newblock Eur. Phys. J. A \textbf{34}, 185--196 (2007).
\newblock \doi{10.1140/epja/i2007-10500-9}

\bibitem{Borasoy:2007vk}
Borasoy, B., Epelbaum, E., Krebs, H., Lee, D., Mei{\ss}ner, U.G.: {Dilute
  neutron matter on the lattice at next-to-leading order in chiral effective
  field theory}.
\newblock Eur. Phys. J. A \textbf{35}, 357--367 (2008).
\newblock \doi{10.1140/epja/i2008-10545-2}

\bibitem{Borasoy:2005yc}
Borasoy, B., Krebs, H., Lee, D., Mei{\ss}ner, U.G.: {The Triton and
  three-nucleon force in nuclear lattice simulations}.
\newblock Nucl. Phys. A \textbf{768}, 179--193 (2006).
\newblock \doi{10.1016/j.nuclphysa.2006.01.009}

\bibitem{Bour:2011ef}
Bour, S., Koenig, S., Lee, D., Hammer, H.W., Mei{\ss}ner, U.G.: {Topological
  phases for bound states moving in a finite volume}.
\newblock Phys. Rev. D \textbf{84}, 091,503 (2011).
\newblock \doi{10.1103/PhysRevD.84.091503}

\bibitem{Bour:2014bxa}
Bour, S., Lee, D., Hammer, H.W., Mei{\ss}ner, U.G.: {Ab initio Lattice Results
  for Fermi Polarons in Two Dimensions}.
\newblock Phys. Rev. Lett. \textbf{115}(18), 185,301 (2015).
\newblock \doi{10.1103/PhysRevLett.115.185301}

\bibitem{Bovermann:2019jbt}
Bovermann, L., Epelbaum, E., Krebs, H., Lee, D.: {Scattering phase shifts and
  mixing angles for an arbitrary number of coupled channels on the lattice}.
\newblock Phys. Rev. C \textbf{100}(6), 064,001 (2019).
\newblock \doi{10.1103/PhysRevC.100.064001}

\bibitem{Brockmann:1992in}
Brockmann, R., Frank, J.: goo.
\newblock Phys. Rev. Lett. \textbf{68}, 1830--1833 (1992)

\bibitem{Bulgac:2005pj}
Bulgac, A., Drut, J.E., Magierski, P.: {Spin 1/2 Fermions in the unitary
  regime: A Superfluid of a new type}.
\newblock Phys. Rev. Lett. \textbf{96}, 090,404 (2006).
\newblock \doi{10.1103/PhysRevLett.96.090404}

\bibitem{Burovski:2006}
{Burovski}, E., {Prokof'ev}, N., {Svistunov}, B., {Troyer}, M.: {The Fermi
  Hubbard model at unitarity}.
\newblock New Journal of Physics \textbf{8}(8), 153 (2006).
\newblock \doi{10.1088/1367-2630/8/8/153}

\bibitem{CalleCordon:2009ps}
Calle~Cordon, A., Ruiz~Arriola, E.: {Serber symmetry, Large N(c) and
  Yukawa-like One Boson Exchange Potentials}.
\newblock Phys. Rev. C \textbf{80}, 014,002 (2009).
\newblock \doi{10.1103/PhysRevC.80.014002}

\bibitem{Chandrasekharan:2003wy}
Chandrasekharan, S., Pepe, M., Steffen, F.D., Wiese, U.J.: {Nonlinear
  realization of chiral symmetry on the lattice}.
\newblock JHEP \textbf{12}, 035 (2003).
\newblock \doi{10.1088/1126-6708/2003/12/035}

\bibitem{Chen:2003vy}
Chen, J.W., Kaplan, D.B.: {A Lattice theory for low-energy fermions at finite
  chemical potential}.
\newblock Phys. Rev. Lett. \textbf{92}, 257,002 (2004).
\newblock \doi{10.1103/PhysRevLett.92.257002}

\bibitem{Chen:2004rq}
Chen, J.W., Lee, D., Sch\"afer, T.: {Inequalities for light nuclei in the
  Wigner symmetry limit}.
\newblock Phys. Rev. Lett. \textbf{93}, 242,302 (2004).
\newblock \doi{10.1103/PhysRevLett.93.242302}

\bibitem{Davoudi:2011md}
Davoudi, Z., Savage, M.J.: {Improving the Volume Dependence of Two-Body Binding
  Energies Calculated with Lattice QCD}.
\newblock Phys. Rev. D \textbf{84}, 114,502 (2011).
\newblock \doi{10.1103/PhysRevD.84.114502}

\bibitem{Demol:2019yjt}
Demol, P., Duguet, T., Ekstr\"om, A., Frosini, M., Hebeler, K., K\"onig, S.,
  Lee, D., Schwenk, A., Som\`a, V., Tichai, A.: {Improved many-body expansions
  from eigenvector continuation}.
\newblock Phys. Rev. C \textbf{101}(4), 041,302 (2020).
\newblock \doi{10.1103/PhysRevC.101.041302}

\bibitem{Demol:2020mzd}
Demol, P., Frosini, M., Tichai, A., Som\`a, V., Duguet, T.: {Bogoliubov
  many-body perturbation theory under constraint}.
\newblock Annals Phys. \textbf{424}, 168,358 (2021).
\newblock \doi{10.1016/j.aop.2020.168358}

\bibitem{Drut:2012a}
{Drut}, J.E., {Nicholson}, A.N.: {Lattice methods for strongly interacting
  many-body systems}.
\newblock J. Phys. G: Nucl. Part. Phys. \textbf{40}(4), 043101 (2013).
\newblock \doi{10.1088/0954-3899/40/4/043101}

\bibitem{Elhatisari:2017eno}
Elhatisari, S., Epelbaum, E., Krebs, H., L\"ahde, T.A., Lee, D., Li, N., Lu,
  B.n., Mei{\ss}ner, U.G., Rupak, G.: {Ab initio Calculations of the Isotopic
  Dependence of Nuclear Clustering}.
\newblock Phys. Rev. Lett. \textbf{119}(22), 222,505 (2017).
\newblock \doi{10.1103/PhysRevLett.119.222505}

\bibitem{Elhatisari:2016hby}
Elhatisari, S., Lee, D., Mei{\ss}ner, U.G., Rupak, G.: {Nucleon-deuteron
  scattering using the adiabatic projection method}.
\newblock Eur. Phys. J. A \textbf{52}(6), 174 (2016).
\newblock \doi{10.1140/epja/i2016-16174-2}

\bibitem{Elhatisari:2015iga}
Elhatisari, S., Lee, D., Rupak, G., Epelbaum, E., Krebs, H., L\"ahde, T.A.,
  Luu, T., Mei{\ss}ner, U.G.: {Ab initio alpha-alpha scattering}.
\newblock Nature \textbf{528}, 111 (2015).
\newblock \doi{10.1038/nature16067}

\bibitem{Elhatisari:2016owd}
Elhatisari, S., et~al.: {Nuclear binding near a quantum phase transition}.
\newblock Phys. Rev. Lett. \textbf{117}(13), 132,501 (2016).
\newblock \doi{10.1103/PhysRevLett.117.132501}

\bibitem{Epelbaum:2012qn}
Epelbaum, E., Krebs, H., Lahde, T.A., Lee, D., Mei{\ss}ner, U.G.: {Structure
  and rotations of the Hoyle state}.
\newblock Phys. Rev. Lett. \textbf{109}, 252,501 (2012).
\newblock \doi{10.1103/PhysRevLett.109.252501}

\bibitem{Epelbaum:2012iu}
Epelbaum, E., Krebs, H., L\"ahde, T.A., Lee, D., Mei{\ss}ner, U.G.: {Viability
  of Carbon-Based Life as a Function of the Light Quark Mass}.
\newblock Phys. Rev. Lett. \textbf{110}(11), 112,502 (2013).
\newblock \doi{10.1103/PhysRevLett.110.112502}

\bibitem{Epelbaum:2013paa}
Epelbaum, E., Krebs, H., L\"ahde, T.A., Lee, D., Mei{\ss}ner, U.G., Rupak, G.:
  {Ab Initio Calculation of the Spectrum and Structure of $^{16}$O}.
\newblock Phys. Rev. Lett. \textbf{112}(10), 102,501 (2014).
\newblock \doi{10.1103/PhysRevLett.112.102501}

\bibitem{Epelbaum:2009rkz}
Epelbaum, E., Krebs, H., Lee, D., Mei{\ss}ner, U.G.: {Ground state energy of
  dilute neutron matter at next-to-leading order in lattice chiral effective
  field theory}.
\newblock Eur. Phys. J. A \textbf{40}, 199--213 (2009).
\newblock \doi{10.1140/epja/i2009-10755-0}

\bibitem{Epelbaum:2009zsa}
Epelbaum, E., Krebs, H., Lee, D., Mei{\ss}ner, U.G.: {Lattice chiral effective
  field theory with three-body interactions at next-to-next-to-leading order}.
\newblock Eur. Phys. J. A \textbf{41}, 125--139 (2009).
\newblock \doi{10.1140/epja/i2009-10764-y}

\bibitem{Epelbaum:2010xt}
Epelbaum, E., Krebs, H., Lee, D., Mei{\ss}ner, U.G.: {Lattice calculations for
  A=3,4,6,12 nuclei using chiral effective field theory}.
\newblock Eur. Phys. J. A \textbf{45}, 335--352 (2010).
\newblock \doi{10.1140/epja/i2010-11009-x}

\bibitem{Epelbaum:2009pd}
Epelbaum, E., Krebs, H., Lee, D., Mei{\ss}ner, U.G.: {Lattice effective field
  theory calculations for A = 3,4,6,12 nuclei}.
\newblock Phys. Rev. Lett. \textbf{104}, 142,501 (2010).
\newblock \doi{10.1103/PhysRevLett.104.142501}

\bibitem{Epelbaum:2011md}
Epelbaum, E., Krebs, H., Lee, D., Mei{\ss}ner, U.G.: {Ab initio calculation of
  the Hoyle state}.
\newblock Phys. Rev. Lett. \textbf{106}, 192,501 (2011).
\newblock \doi{10.1103/PhysRevLett.106.192501}

\bibitem{Frame:2017fah}
Frame, D., He, R., Ipsen, I., Lee, D., Lee, D., Rrapaj, E.: {Eigenvector
  continuation with subspace learning}.
\newblock Phys. Rev. Lett. \textbf{121}(3), 032,501 (2018).
\newblock \doi{10.1103/PhysRevLett.121.032501}

\bibitem{Frame:2020mvv}
Frame, D., L\"ahde, T.A., Lee, D., Mei{\ss}ner, U.G.: {Impurity Lattice Monte
  Carlo for Hypernuclei}.
\newblock Eur. Phys. J. A \textbf{56}(10), 248 (2020).
\newblock \doi{10.1140/epja/s10050-020-00257-y}

\bibitem{Frame:2019jsw}
Frame, D.K.: {Ab Initio Simulations of Light Nuclear Systems Using Eigenvector
  Continuation and Auxiliary Field Monte Carlo}.
\newblock Other thesis (2019)

\bibitem{He:2019ipt}
He, R., Li, N., Lu, B.N., Lee, D.: {Superfluid Condensate Fraction and Pairing
  Wave Function of the Unitary Fermi Gas}.
\newblock Phys. Rev. A \textbf{101}(6), 063,615 (2020).
\newblock \doi{10.1103/PhysRevA.101.063615}

\bibitem{Hoffman:2016jqv}
Hoffman, M.D., Loheac, A.C., Porter, W.J., Drut, J.E.: {Thermodynamics of
  one-dimensional SU(4) and SU(6) fermions with attractive interactions}.
\newblock Phys. Rev. A \textbf{95}(3), 033,602 (2017).
\newblock \doi{10.1103/PhysRevA.95.033602}

\bibitem{Hubbard:1959ub}
Hubbard, J.: {Calculation of partition functions}.
\newblock Phys. Rev. Lett. \textbf{3}, 77--80 (1959).
\newblock \doi{10.1103/PhysRevLett.3.77}

\bibitem{Kanada-Enyo:2020zzf}
Kanada-En'yo, Y., Lee, D.: {Effective interactions between nuclear clusters}.
\newblock Phys. Rev. C \textbf{103}(2), 024,318 (2021).
\newblock \doi{10.1103/PhysRevC.103.024318}

\bibitem{Kaplan:1996rk}
Kaplan, D.B., Manohar, A.V.: {The Nucleon-nucleon potential in the 1/N(c)
  expansion}.
\newblock Phys. Rev. C \textbf{56}, 76--83 (1997).
\newblock \doi{10.1103/PhysRevC.56.76}

\bibitem{Kaplan:1995yg}
Kaplan, D.B., Savage, M.J.: {The Spin flavor dependence of nuclear forces from
  large n QCD}.
\newblock Phys. Lett. B \textbf{365}, 244--251 (1996).
\newblock \doi{10.1016/0370-2693(95)01277-X}

\bibitem{Koenig:2011vaa}
Koenig, S., Lee, D., Hammer, H.W.: {Non-relativistic bound states in a finite
  volume}.
\newblock Annals Phys. \textbf{327}, 1450--1471 (2012).
\newblock \doi{10.1016/j.aop.2011.12.015}

\bibitem{Konig:2017krd}
K\"onig, S., Lee, D.: {Volume Dependence of N-Body Bound States}.
\newblock Phys. Lett. B \textbf{779}, 9--15 (2018).
\newblock \doi{10.1016/j.physletb.2018.01.060}

\bibitem{Koonin:1986}
Koonin, S.E.: {Auxiliary-field Monte Carlo methods}.
\newblock Journal of Statistical Physics \textbf{43}(5-6), 985--990 (1986).
\newblock \doi{10.1007/BF02628325}

\bibitem{Korber:2017emn}
K\"orber, C., Berkowitz, E., Luu, T.: {Sampling General N-Body Interactions
  with Auxiliary Fields}.
\newblock EPL \textbf{119}(6), 60,006 (2017).
\newblock \doi{10.1209/0295-5075/119/60006}

\bibitem{Korber:2015rce}
K\"orber, C., Luu, T.: {Applying Twisted Boundary Conditions for Few-body
  Nuclear Systems}.
\newblock Phys. Rev. C \textbf{93}(5), 054,002 (2016).
\newblock \doi{10.1103/PhysRevC.93.054002}

\bibitem{Lahde:2013uqa}
L\"ahde, T.A., Epelbaum, E., Krebs, H., Lee, D., Mei{\ss}ner, U.G., Rupak, G.:
  {Lattice Effective Field Theory for Medium-Mass Nuclei}.
\newblock Phys. Lett. B \textbf{732}, 110--115 (2014).
\newblock \doi{10.1016/j.physletb.2014.03.023}

\bibitem{Lahde:2015ona}
L\"ahde, T.A., Luu, T., Lee, D., Mei{\ss}ner, U.G., Epelbaum, E., Krebs, H.,
  Rupak, G.: {Nuclear Lattice Simulations using Symmetry-Sign Extrapolation}.
\newblock Eur. Phys. J. A \textbf{51}(7), 92 (2015).
\newblock \doi{10.1140/epja/i2015-15092-1}

\bibitem{Lahde:2019npb}
L\"ahde, T.A., Mei{\ss}ner, U.G.: {Nuclear Lattice Effective Field Theory}: {An
  introduction}, vol. 957.
\newblock Springer (2019).
\newblock \doi{10.1007/978-3-030-14189-9}

\bibitem{Lee:2004ze}
Lee, D.: {Inequalities for low-energy symmetric nuclear matter}.
\newblock Phys. Rev. C \textbf{70}, 064,002 (2004).
\newblock \doi{10.1103/PhysRevC.70.064002}

\bibitem{Lee:2004hc}
Lee, D.: {Pressure inequalities for nuclear and neutron matter}.
\newblock Phys. Rev. C \textbf{71}, 044,001 (2005).
\newblock \doi{10.1103/PhysRevC.71.044001}

\bibitem{Lee:2007eu}
Lee, D.: {Spectral convexity for attractive SU(2N) fermions}.
\newblock Phys. Rev. Lett. \textbf{98}, 182,501 (2007).
\newblock \doi{10.1103/PhysRevLett.98.182501}

\bibitem{Lee:2008fa}
Lee, D.: {Lattice simulations for few- and many-body systems}.
\newblock Prog. Part. Nucl. Phys. \textbf{63}, 117--154 (2009).
\newblock \doi{10.1016/j.ppnp.2008.12.001}

\bibitem{Lee:2004si}
Lee, D., Borasoy, B., Sch\"afer, T.: {Nuclear lattice simulations with chiral
  effective field theory}.
\newblock Phys. Rev. C \textbf{70}, 014,007 (2004).
\newblock \doi{10.1103/PhysRevC.70.014007}

\bibitem{Lee:2004qd}
Lee, D., Sch\"afer, T.: {Neutron matter on the lattice with pionless effective
  field theory}.
\newblock Phys. Rev. C \textbf{72}, 024,006 (2005).
\newblock \doi{10.1103/PhysRevC.72.024006}

\bibitem{Lee:2005is}
Lee, D., Sch\"afer, T.: {Cold dilute neutron matter on the lattice. I. Lattice
  virial coefficients and large scattering lengths}.
\newblock Phys. Rev. C \textbf{73}, 015,201 (2006).
\newblock \doi{10.1103/PhysRevC.73.015201}

\bibitem{Lee:2005it}
Lee, D., Sch\"afer, T.: {Cold dilute neutron matter on the lattice. II. Results
  in the unitary limit}.
\newblock Phys. Rev. C \textbf{73}, 015,202 (2006).
\newblock \doi{10.1103/PhysRevC.73.015202}

\bibitem{Lee:2020esp}
Lee, D., et~al.: {Hidden Spin-Isospin Exchange Symmetry}.
\newblock Phys. Rev. Lett. \textbf{127}(6), 062,501 (2021).
\newblock \doi{10.1103/PhysRevLett.127.062501}

\bibitem{Lewis:2000cc}
Lewis, R., Ouimet, P.P.A.: {Lattice regularization for chiral perturbation
  theory}.
\newblock Phys. Rev. D \textbf{64}, 034,005 (2001).
\newblock \doi{10.1103/PhysRevD.64.034005}

\bibitem{Li:2019ldq}
Li, N., Elhatisari, S., Epelbaum, E., Lee, D., Lu, B., Mei{\ss}ner, U.G.:
  {Galilean invariance restoration on the lattice}.
\newblock Phys. Rev. C \textbf{99}(6), 064,001 (2019).
\newblock \doi{10.1103/PhysRevC.99.064001}

\bibitem{Li:2018ymw}
Li, N., Elhatisari, S., Epelbaum, E., Lee, D., Lu, B.N., Mei{\ss}ner, U.G.:
  {Neutron-proton scattering with lattice chiral effective field theory at
  next-to-next-to-next-to-leading order}.
\newblock Phys. Rev. C \textbf{98}(4), 044,002 (2018).
\newblock \doi{10.1103/PhysRevC.98.044002}

\bibitem{Lu:2014xfa}
Lu, B.N., L\"ahde, T.A., Lee, D., Mei{\ss}ner, U.G.: {Breaking and restoration
  of rotational symmetry on the lattice for bound state multiplets}.
\newblock Phys. Rev. D \textbf{90}(3), 034,507 (2014).
\newblock \doi{10.1103/PhysRevD.90.034507}

\bibitem{Lu:2015riz}
Lu, B.N., L\"ahde, T.A., Lee, D., Mei{\ss}ner, U.G.: {Precise determination of
  lattice phase shifts and mixing angles}.
\newblock Phys. Lett. B \textbf{760}, 309--313 (2016).
\newblock \doi{10.1016/j.physletb.2016.06.081}

\bibitem{Lu:2019nbg}
Lu, B.N., Li, N., Elhatisari, S., Lee, D., Drut, J.E., L\"ahde, T.A., Epelbaum,
  E., Mei{\ss}ner, U.G.: {$Ab Initio$ Nuclear Thermodynamics}.
\newblock Phys. Rev. Lett. \textbf{125}(19), 192,502 (2020).
\newblock \doi{10.1103/PhysRevLett.125.192502}

\bibitem{Lu:2018bat}
Lu, B.N., Li, N., Elhatisari, S., Lee, D., Epelbaum, E., Mei{\ss}ner, U.G.:
  {Essential elements for nuclear binding}.
\newblock Phys. Lett. B \textbf{797}, 134,863 (2019).
\newblock \doi{10.1016/j.physletb.2019.134863}

\bibitem{Luscher:1986pf}
Luscher, M.: {Volume Dependence of the Energy Spectrum in Massive Quantum Field
  Theories. 2. Scattering States}.
\newblock Commun. Math. Phys. \textbf{105}, 153--188 (1986).
\newblock \doi{10.1007/BF01211097}

\bibitem{Muller:1999cp}
Muller, H.M., Koonin, S.E., Seki, R., van Kolck, U.: {Nuclear matter on a
  lattice}.
\newblock Phys. Rev. C \textbf{61}, 044,320 (2000).
\newblock \doi{10.1103/PhysRevC.61.044320}

\bibitem{Pine:2013zja}
Pine, M., Lee, D., Rupak, G.: {Adiabatic projection method for scattering and
  reactions on the lattice}.
\newblock Eur. Phys. J. A \textbf{49}, 151 (2013).
\newblock \doi{10.1140/epja/i2013-13151-3}

\bibitem{Rokash:2016tqh}
Rokash, A., Epelbaum, E., Krebs, H., Lee, D.: {Effective forces between quantum
  bound states}.
\newblock Phys. Rev. Lett. \textbf{118}(23), 232,502 (2017).
\newblock \doi{10.1103/PhysRevLett.118.232502}

\bibitem{Rokash:2015hra}
Rokash, A., Pine, M., Elhatisari, S., Lee, D., Epelbaum, E., Krebs, H.:
  {Scattering cluster wave functions on the lattice using the adiabatic
  projection method}.
\newblock Phys. Rev. C \textbf{92}(5), 054,612 (2015).
\newblock \doi{10.1103/PhysRevC.92.054612}

\bibitem{RuizArriola:2016vap}
Ruiz~Arriola, E.: {Low Scale Saturation of Effective NN Interactions and Their
  Symmetries}.
\newblock Symmetry \textbf{8}(6), 42 (2016).
\newblock \doi{10.3390/sym8060042}

\bibitem{Rupak:2013aue}
Rupak, G., Lee, D.: {Radiative capture reactions in lattice effective field
  theory}.
\newblock Phys. Rev. Lett. \textbf{111}(3), 032,502 (2013).
\newblock \doi{10.1103/PhysRevLett.111.032502}

\bibitem{Sarkar:2020mad}
Sarkar, A., Lee, D.: {Convergence of Eigenvector Continuation}.
\newblock Phys. Rev. Lett. \textbf{126}(3), 032,501 (2021).
\newblock \doi{10.1103/PhysRevLett.126.032501}

\bibitem{Shen:2021kqr}
Shen, S., L\"ahde, T.A., Lee, D., Mei{\ss}ner, U.G.: {Wigner SU(4) symmetry,
  clustering, and the spectrum of $^{12}$C}  (2021)

\bibitem{Shushpanov:1998ms}
Shushpanov, I.A., Smilga, A.V.: {Chiral perturbation theory with lattice
  regularization}.
\newblock Phys. Rev. D \textbf{59}, 054,013 (1999).
\newblock \doi{10.1103/PhysRevD.59.054013}

\bibitem{Song:2021yst}
Song, Y.H., Kim, Y., Li, N., Lu, B.N., He, R., Lee, D.: {Quantum Many-Body
  Calculations using Body-Centered Cubic Lattices}  (2021)

\bibitem{Stellin:2018fkj}
Stellin, G., Elhatisari, S., Mei{\ss}ner, U.G.: {Breaking and restoration of
  rotational symmetry in the low-energy spectrum of light alpha-conjugate
  nuclei on the lattice I: $^{8}\mathrm{Be}$ and $^{12}\mathrm{C}$}.
\newblock Eur. Phys. J. A \textbf{54}(12), 232 (2018).
\newblock \doi{10.1140/epja/i2018-12671-6}

\bibitem{Stratonovich:1958}
Stratonovich, R.L.: On a method of calculating quantum distribution functions.
\newblock Soviet Phys. Doklady \textbf{2}, 416--419 (1958)

\bibitem{Timoteo:2011tt}
Timoteo, V., Szpigel, S., Ruiz~Arriola, E.: {Symmetries of the Similarity
  Renormalization Group for Nuclear Forces}.
\newblock Phys. Rev. C \textbf{86}, 034,002 (2012).
\newblock \doi{10.1103/PhysRevC.86.034002}

\bibitem{Weinberg:1990rz}
Weinberg, S.: Nuclear forces from chiral lagrangians.
\newblock Phys. Lett. \textbf{B251}, 288--292 (1990)

\bibitem{Weinberg:1991um}
Weinberg, S.: Effective chiral lagrangians for nucleon - pion interactions and
  nuclear forces.
\newblock Nucl. Phys. \textbf{B363}, 3--18 (1991)

\bibitem{Weinberg:1992yk}
Weinberg, S.: Three body interactions among nucleons and pions.
\newblock Phys. Lett. \textbf{B295}, 114--121 (1992)

\bibitem{Wigner:1936dx}
Wigner, E.: {On the Consequences of the Symmetry of the Nuclear Hamiltonian on
  the Spectroscopy of Nuclei}.
\newblock Phys. Rev. \textbf{51}, 106--119 (1937).
\newblock \doi{10.1103/PhysRev.51.106}

\bibitem{Wingate:2006wy}
Wingate, M.: {Field theoretic study of a cold Fermi gas in the unitary limit}.
\newblock PoS \textbf{LAT2006}, 153 (2006).
\newblock \doi{10.22323/1.032.0153}

\bibitem{Wlazlowski:2014jna}
Wlaz{\l}owski, G., Holt, J.W., Moroz, S., Bulgac, A., Roche, K.J.:
  {Auxiliary-Field Quantum Monte Carlo Simulations of Neutron Matter in Chiral
  Effective Field Theory}.
\newblock Phys. Rev. Lett. \textbf{113}(18), 182,503 (2014).
\newblock \doi{10.1103/PhysRevLett.113.182503}

\end{thebibliography}

\end{document}